\documentclass[smallextended]{svjour3}
\usepackage{graphicx, amsmath, amssymb}
\usepackage[caption=false]{subfig}

\newcommand{\ket}[1]{{\left| #1 \right\rangle}}
\newcommand{\ketbra}[2]{{\left| #1 \middle\rangle \middle \langle #2 \right|}}

\journalname{Quantum Inf Process}

\begin{document}

\title{Quantum Walk on the Line through Potential Barriers}

\author{Thomas G.~Wong}

\authorrunning{T.~G.~Wong}

\institute{T.~G.~Wong \at
	   Faculty of Computing, University of Latvia, Rai\c{n}a bulv.~19, R\=\i ga, LV-1586, Latvia \\
	   \email{twong@lu.lv}
}

\date{Received: date / Accepted: date}

\maketitle

\begin{abstract}
	Quantum walks are well-known for their ballistic dispersion, traveling $\mathrm{\Theta}(t)$ away in $t$ steps, which is quadratically faster than a classical random walk's diffusive spreading. In physical implementations of the walk, however, the particle may need to tunnel through a potential barrier to hop, and a naive calculation suggests this could eliminate the ballistic transport. We show by explicit calculation, however, that such a loss does not occur. Rather, the $\mathrm{\Theta}(t)$ dispersion is retained, with only the coefficient changing, which additionally gives a way to detect and quantify the hopping errors in experiments.
\end{abstract}

%-------------------------------------------------------------------------------
% Main Matter
%-------------------------------------------------------------------------------

\section{Introduction}

Quantum walks are the quantum analogues of classical random walks. In the same way that classical random walks, or Markov chains, serve as useful algorithmic tools for classical computing \cite{Norris1998}, quantum walks are the basis for several important quantum algorithms, including search \cite{SKW2003,CG2004}, element distinctness \cite{Ambainis2004}, and boolean formula evaluation \cite{FGG2008,Ambainis2010}. For a comprehensive review of quantum walks, see \cite{Venegas2012}.

Before quantum walks were applied algorithmically, however, their basic properties were investigated, beginning with how they evolve on the one-dimensional infinite lattice. This began with Meyer \cite{Meyer1996a} in the context of quantum cellular automata, who showed that a discrete-time quantum walk necessitates an internal degree of freedom to evolve non-trivially. Interpreting this internal degree of freedom as spin, he showed that the evolution's continuum limit is the Dirac equation of a spin-1/2 particle. Later, this work was translated into the language of quantum walks and extended by Ambainis \textit{et al.~}\cite{Ambainis2001,NV2000}, revealing an important distinction between classical and quantum walks: quantum walks spread quadratically faster than classical ones. This was also shown to be true for continuous-time quantum walks \cite{Childs2003}, which do not require the internal degree of freedom \cite{FG1998b}. Since these investigations, there has been continued interest in one-dimensional quantum walks; just some examples include \cite{Konno2002,GJS2004,benAvraham2004,Konno2005,Gottlieb2005,SA2013}.

Spurred by this substantial theoretical interest, several experimental groups have implemented one-dimensional quantum walks in a variety of physical systems, including linear optical resonators \cite{Bouwmeester1999}, nuclear magnetic resonance samples \cite{Ryan2005}, photons in waveguide lattices \cite{Perets2008,Schreiber2010}, photons in interferometric networks \cite{Broome2010}, optically trapped atoms \cite{Karski2009}, and linearly trapped ions \cite{Schmitz2009,Matjeschk2012,Zahringer2010}. For most of these, the walk occurred in discrete-time.

In this paper, we also focus on the discrete-time quantum walk on the one-dimensional infinite lattice, but with two important differences that we will introduce below. To begin, we use computational basis states labeled by the integers $\{ \dots, \ket{-2}, \ket{-1}, \ket{0}, \ket{1}, \ket{2}, \dots \}$ for the position of the particle. Since there are two possible directions for the particle to hop (left and right), we take the particle to be a two-component spinor with computational basis states $\{ \ket{\leftarrow}, \ket{\rightarrow} \}$. Then the state of the system is a vector in $\mathbb{C}^\infty \otimes \mathbb{C}^2$, where the first tensor factor corresponds to the position of the particle, and the second to its internal spin or ``coin'' state. The quantum walk is given by repeated applications of the unitary operator
\[ S (I \otimes H), \]
where $H$ is the ``Hadamard coin,'' which operates on the coin space and takes
\begin{align*}
	H \ket{\leftarrow} &= \frac{1}{\sqrt{2}} \left( \ket{\leftarrow} + \ket{\rightarrow} \right) \\
	H \ket{\rightarrow} &= \frac{1}{\sqrt{2}} \left( \ket{\leftarrow} - \ket{\rightarrow} \right),
\end{align*}
and $S$ is the ``flip-flop'' shift
\begin{align*}
	S \ket{i,\leftarrow} &= \ket{i-1,\rightarrow} \\
	S \ket{i,\rightarrow} &= \ket{i+1,\leftarrow} ,
\end{align*}
which flips the internal state of the walker when hopping. The use of the flip-flop shift $S$ is the first important distinction of our analysis, which contrasts with typical investigations of 1D quantum walks that instead use the ``moving'' shift, where the direction of the walker is unchanged after the hop. There are several reasons for our choice of the flip-flop shift. First, it is the natural operator for a quantum walk using a macroscopic Bose-Einstein condensate \cite{Chandrashekar2006}, where the condensate atoms shift by exchanging photons with two laser fields, undergoing a stimulated Raman transition. This imparts a momentum kick to the atoms that physically shifts them while flipping their internal states. A second reason for using the flip-flop shift is that it has not been previously analyzed for the 1D quantum walk. A third reason is that the flip-flop shift is Hermitian, which will be important in a moment when we introduce potential barriers. A fourth reason for using the flip-flop shift is algorithmic: it allows quantum walks to search graphs quickly, whereas the moving shift, by contrast, yields a slow search algorithm \cite{AKR2005}. Furthermore, as a fifth reason, defining the flip-flop shift on non-Euclidian lattices, such as the complete graph \cite{Wong2015b} or hypercube \cite{SKW2003}, is natural, whereas it is unclear what the moving shift would be. Finally, a sixth reason is that for a quantum walk alone (\textit{i.e.}, simply applying the coin and shift operator, so no oracle or algorithmic modifications), the flip-flop shift yields the same probability distribution as the moving shift, apart from a directional flip.

\begin{figure}
\begin{center}
	\includegraphics{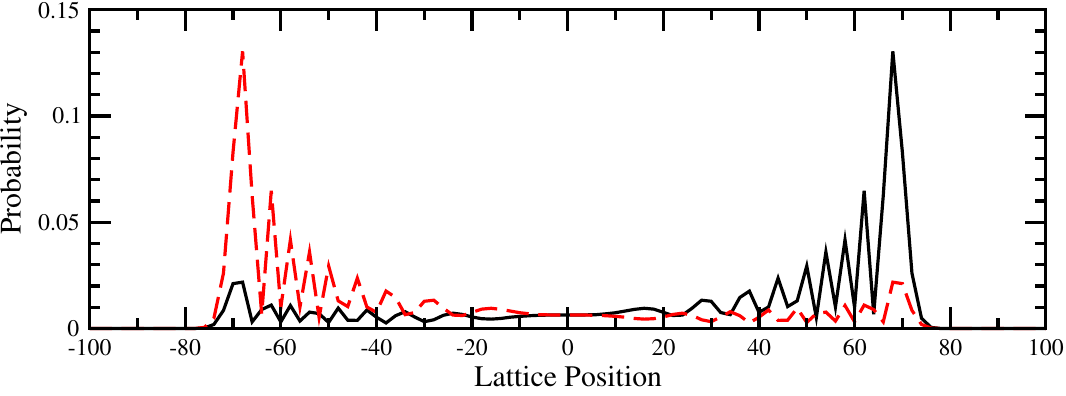}
	\caption{\label{fig:flip} Probability at even lattice sites for an initially left-moving particle localized at 0 after 100 steps of the quantum walk. The solid black curve is with the flip-flop shift, and the dashed red curve is with the moving shift. Note the probability at odd sites is zero.}
\end{center}
\end{figure}
	
To illustrate this final point, say the quantum walk starts in the middle of the lattice (\textit{i.e.}, at position 0), pointing left:
\begin{equation}
	\label{eq:initial}
	\ket{\psi_0} = \ket{0} \otimes \ket{\leftarrow}.
\end{equation}
The probability distribution of the quantum walk after 100 steps, for both the flip-flop and moving shifts, is shown in Figure~\ref{fig:flip}. Surprising, the flip-flop shift causes the walk to become right-moving, while the moving shift retains the initial left-moving behavior. In addition, the probability distributions are exact reflection of each other. To see why this occurs, note that after two steps of the quantum walk, the state of the system with each of the shifts is
\begin{align*}
	\ket{\psi_{2,\text{flip-flop}}} &= \frac{1}{2} \Big[ \ket{-2} \otimes \ket{\rightarrow} - \ket{0} \otimes \big( \ket{\leftarrow} - \ket{\rightarrow} \big) - \ket{2} \otimes \ket{\leftarrow} \Big] \\
	\ket{\psi_{2,\text{moving}}} &= \frac{1}{2} \Big[ \ket{-2} \otimes \ket{\leftarrow} + \ket{0} \otimes \big( \ket{\leftarrow} + \ket{\rightarrow} \big) - \ket{2} \otimes \ket{\rightarrow} \Big] .
\end{align*}
At this point, the probability distributions of the walks are identical. It is during the third step that the directional flip occurs. To see this, focus on the coin state at the middle vertex $\ket{0}$, which for the flip-flop shift is $(\ket{\leftarrow} - \ket{\rightarrow})/\sqrt{2}$, and for the moving shift is $(\ket{\leftarrow} + \ket{\rightarrow})/\sqrt{2}$. When the Hadamard transform acts on the flip-flop shift's middle coin state, it becomes $\ket{\rightarrow}$, so the walk has become right-moving. For the rest of the walk, it continues to be right-moving, with the probability distribution mirroring the moving shift's left-moving walk. When it acts on the flip-flop shift's coin state, however, it becomes $\ket{\leftarrow}$, so it retains its left-moving bias. 

\begin{figure}
\begin{center}
	\includegraphics{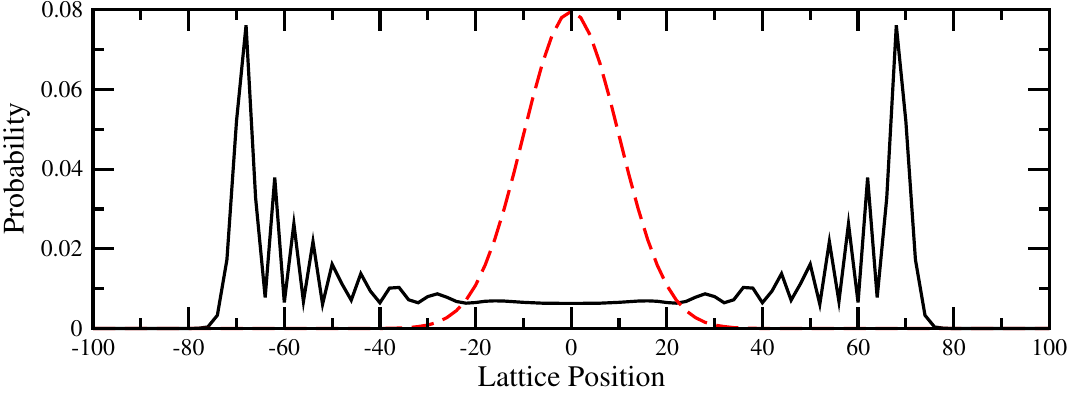}
	\caption{\label{fig:quantum_classical} Probability at even lattice sites after 100 steps of the quantum walk (solid black) and classical random walk (dashed red) with an unbiased initial state in the middle of the lattice (so the quantum starts in \eqref{eq:unbiased}, and the classical starts with probability $1$ at the middle). Note the probability at odd sites is zero.}
\end{center}
\end{figure}

Since the quantum walk is a purely real operator, if the system starts the unbiased initial state localized in the middle of the lattice,
\begin{equation}
	\label{eq:unbiased}
	\ket{\psi_0} = \ket{0} \otimes \frac{1}{\sqrt{2}} \left( \ket{\leftarrow} + i \ket{\rightarrow} \right),
\end{equation}
then the flip-flop and moving shifts will yield exactly the same probability distribution. This distribution after 100 steps of the walk is shown in Figure~\ref{fig:quantum_classical}, along with the classical random walk's probability distribution; this reveals the quantum walk's linear, ballistic propagation, which is quadratically faster than the classical walk's square-root diffusion. That is, after $t$ steps, the quantum walk spreads an average distance $\mathrm{\Theta}(t)$ while the classical random walk spreads $\mathrm{\Theta}(\sqrt{t})$. Note since the quantum walk will not mix real and imaginary components, rather than using the unbiased initial state \eqref{eq:unbiased}, it suffices to analyze the left-moving initial state \eqref{eq:initial} instead, and we assume this for the rest of the paper.

\begin{figure}
\begin{center}
	\setcounter{subfigure}{-1}
	\subfloat{
		\includegraphics{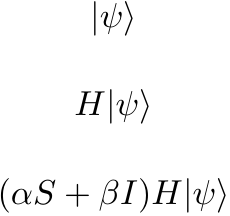}
	} \quad
	\subfloat[]{
		\includegraphics{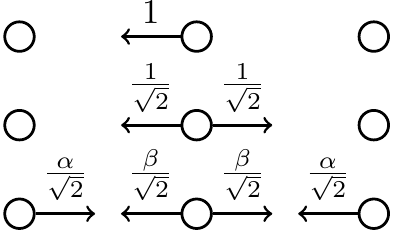}
		\label{fig:simplex_diagram_H}
	} \quad \quad
	\subfloat[]{
		\includegraphics{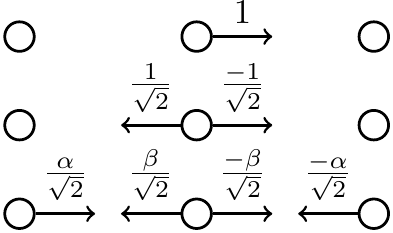}
		\label{fig:simplex_diagram_stage1}
	}

	\caption{\label{fig:step} One step of the quantum walk with potential barriers (\textit{i.e.}, the Hadamard coin followed by a shift/identity) for a (a) left-moving and (b) right-moving particle.}
\end{center}
\end{figure}

Rather than just analyzing this 1D quantum walk, we augment it by introducing potential barriers that the quantum walker must tunnel through in order to shift \cite{Wong2015c,AmbainisWong2015}. That is, we replace the flip-flop shift with
\[ S \to \alpha S + \beta I, \]
so there is some amplitude $\alpha$ that the particle successfully tunnels, and some (related) amplitude $\beta$ that it fails and stays put. As previously noted, the flip-flop shift $S$ is Hermitian, so the modified shift is unitary provided $|\alpha|^2 + |\beta|^2 = 1$ and $\alpha\beta^* + \beta\alpha^* = 0$. If the moving shift was used instead, then this operator would not be unitary, and would hence not be an quantum operator. With the potential barriers, a complete step of the quantum walk is
\begin{equation}
	\label{eq:U}
	U = (\alpha S + \beta I) (I \otimes H),
\end{equation}
and its action on a left- or right-moving particle is shown in Figure~\ref{fig:step}. Modifying the shift is a second distinction of this paper from previous work on 1D quantum walks.

There are other physical processes where this modified shift can arise. For example, a quantum walk using atoms trapped in an optical lattice \cite{Karski2009} can be shifted by translating the trapping potential. Even if the trapping potential is moved at constant speed, its acceleration is nonzero at the beginning and end of the motion \cite{Mandel2003}, and this acceleration effectively causes the trapping potential to tilt \cite{RSN1997}, resulting in the particles staying put. As another example, a quantum walk using a macroscopic Bose-Einstein condensate \cite{Chandrashekar2006} shifts when a stimulated Raman transition imparts a momentum kick to the atoms. The uncertainty as to whether all the atoms have shifted once can cause a portion of the condensate to stay put.

\begin{figure}
\begin{center}
	\includegraphics{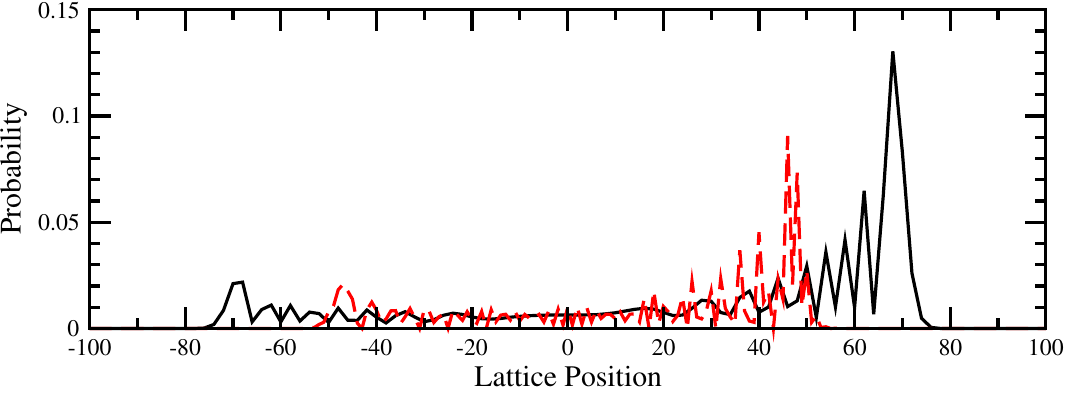}
	\caption{\label{fig:prob_time_barrier} Probability at lattice sites for an initially left-moving particle localized at 0 after 100 steps of the quantum walk. The solid black curve is the barrier-free ($\alpha = 1$ and $\beta = 0$) walk (with only even positions shown), and the dashed red curve is with $\alpha = \cos(0.8)$ and $\beta = i\sin(0.8)$ (with all positions shown).}
\end{center}
\end{figure}

In Figure~\ref{fig:prob_time_barrier}, we plot the probability distribution of the quantum walk after 100 steps both with and without the potential barriers. Recall we are assuming for the rest of the paper that the particle is initially left-moving and localized in the middle of the lattice according to \eqref{eq:initial}, so the particle becomes right-moving. This figure shows that the potential barriers slow down the walk, but how much? As a naive calculation, consider the parameterization $\alpha = \cos\varphi$ and $\beta = i\sin\varphi$ with $\varphi$ small, so $\alpha \approx 1$ and $\beta \approx i\varphi$. Then from Section 4.5.3 of Nielsen and Chuang \cite{NielsenChuang2000}, the error between the perfect shift $S$ and the one with potential barriers $(\alpha S + \beta I)$ is roughly $\varphi$, and after $t$ steps of the quantum walk operator \eqref{eq:U}, the error is upper bounded by $\varphi t$. Since this bound allows the error to increase as the system evolves, one might expect that the characteristic behavior of the quantum walk deteriorates, perhaps also loosing the ballistic propagation.

In the next section, however, we explicitly show that this does not happen. We analytically determine the evolution of the particle and show that the $\mathrm{\Theta}(t)$ dispersion is retained so long as the potential barriers do not asymptotically cause the quantum walk to stop hopping. That is, the potential barriers do not eliminate the characteristic ballistic transport of quantum walks, only scaling it to some degree. Furthermore, this scaling gives a way to detect and quantify if these errors are occurring in experiments. Finally, we end with a comment about continuous-time quantum walks similarly retaining the $\mathrm{\Theta}(t)$ dispersion with potential barriers.

%-------------------------------------------------------------------------------
% Section
%-------------------------------------------------------------------------------

\section{Analysis with Potential Barriers}

Our analysis follows the Fourier method of \cite{Ambainis2001,NV2000}, and we begin by writing the wave function at position $n$ and time $t$ as a two-component spinor
\[ \psi(n,t) = \begin{pmatrix}
	\psi_L(n,t) \\
	\psi_R(n,t) \\
\end{pmatrix}, \]
where $L$ and $R$ denote the left- and right-moving coin components, respectively. Using Figure~\ref{fig:step}, one step of the quantum walk operator \eqref{eq:U} causes this state to become:
\[ \psi(n,t+1) = \begin{pmatrix} \frac{\alpha}{\sqrt{2}} & \frac{-\alpha}{\sqrt{2}} \\ 0 & 0 \end{pmatrix} \psi(n-1,t) + \begin{pmatrix} 0 & 0 \\ \frac{\alpha}{\sqrt{2}} & \frac{\alpha}{\sqrt{2}} \end{pmatrix} \psi(n+1,t) + \begin{pmatrix} \frac{\beta}{\sqrt{2}} & \frac{\beta}{\sqrt{2}} \\ \frac{\beta}{\sqrt{2}} & \frac{-\beta}{\sqrt{2}} \end{pmatrix} \psi(n,t). \]
Let us call these matrices $M_+$, $M_-$, and $M_0$, where the subscript respectively denotes if the position shifts right, left, or stays put. Then
\[ \psi(n,t+1) = M_+ \psi(n-1,t) + M_- \psi(n+1,t) + M_0 \psi(n,t). \]
Now consider the Fourier transform of the position wavefunction, which gives the momentum wavefunction $\tilde{\psi}$  with $k \in [-\pi,\pi]$:
\[ \tilde\psi(k,t) = \sum_n \psi(n,t) e^{ikn}. \]
Then the momentum wavefunction at the next timestep is
\begin{align*}
	\tilde{\psi}(k,t+1) 
		&= \sum_n \psi(n,t+1) e^{ikn} \\
		&= \sum_n \left( M_+ \psi(n-1,t) + M_- \psi(n+1,t) + M_0 \psi(n,t) \right) e^{ikn} \\
		&= \left( M_+ e^{ik} + M_- e^{-ik} + M_0 \right) \sum_n \psi(n,t) e^{ikn} \\
		&= M_k \tilde{\psi}(k,t),
\end{align*}
where we have defined
\[ M_k = M_+ e^{ik} + M_- e^{-ik} + M_0 = \frac{1}{\sqrt{2}} \begin{pmatrix}
	\alpha e^{ik} + \beta & -\alpha e^{ik} + \beta \\
	\alpha e^{-ik} + \beta & \alpha e^{-ik} - \beta \\
\end{pmatrix}. \]
Using this recursive relation, the momentum wavefunction at later time is related to the momentum wavefunction at initial time by
\[ \tilde{\psi}(k,t) = (M_k)^t \tilde{\psi}(k,0), \]
where the initially left-moving particle at the center of the lattice \eqref{eq:initial} has momentum wavefunction
\[ \tilde\psi(k,0) = \begin{pmatrix} 1 \\ 0 \end{pmatrix} \]
for all $k$. So to find the evolution, we diagonalize $M_k$, which makes it easy to take powers of it. To find its eigenvalues $\lambda$, we solve the characteristic equation
\[ 0 = \det(M_k - \lambda I) = \lambda^2 - \alpha \sqrt{2} \cos(k) \lambda + 1. \]
This yields eigenvalues
\[ \lambda_\pm = \frac{\alpha \cos k \pm i \sqrt{2 - \alpha^2 \cos^2k}}{\sqrt{2}} = e^{\pm i \omega_k}, \]
where
\begin{equation}
	\label{eq:omega}
	\cos \omega_k = \frac{\alpha \cos k}{\sqrt{2}}, \quad \sin \omega_k = \frac{\sqrt{2 - \alpha^2 \cos^2k}}{\sqrt{2}}.
\end{equation}
Now let us find the eigenvectors of $M_k$, which satisfy
\[ M_k \begin{pmatrix} u \\ v \end{pmatrix} = e^{\pm i \omega_k} \begin{pmatrix} u \\ v \end{pmatrix}. \]
The top row of this is
\[ \frac{1}{\sqrt{2}} \left[ \left( \alpha e^{ik} + \beta \right) u + \left( -\alpha e^{ik} + \beta \right) v \right] = e^{\pm i \omega_k} u, \]
and with some algebra, it becomes
\[ \frac{\alpha + \beta e^{-ik} - \sqrt{2} e^{i(\pm\omega_k - k)}}{\alpha - \beta e^{-ik} } u = v. \]
Thus the (unnormalized) eigenvectors of $M_k$ are
\[ \begin{pmatrix} u \\ v \end{pmatrix} = \begin{pmatrix} 1 \\ \frac{\alpha + \beta e^{-ik} - \sqrt{2} e^{i(\pm\omega_k - k)}}{\alpha - \beta e^{-ik} } \end{pmatrix}. \]
Let us normalize this by finding the square of its norm:
\begin{align*}
	\left| \begin{pmatrix} u \\ v \end{pmatrix} \right|^2
		&= uu^* + vv^* \\
		&= 1 + \left( \frac{\alpha + \beta e^{-ik} - \sqrt{2} e^{i(\pm\omega_k - k)}}{\alpha - \beta e^{-ik} } \right) \left( \frac{\alpha^* + \beta^* e^{ik} - \sqrt{2} e^{-i(\pm\omega_k - k)}}{\alpha^* - \beta^* e^{ik} } \right)
\end{align*}
Now we assume that $\alpha$ is real and $\beta$ is pure imaginary; for example, they could be parameterized as $\alpha = \cos\varphi$ and $\beta = i\sin\varphi$. Since $|\alpha|^2 + |\beta|^2 = 1$, this implies that $\alpha^2 - \beta^2 = 1$. So the square of the norm of the eigenvectors becomes
\begin{align*}
	\left| \begin{pmatrix} u \\ v \end{pmatrix} \right|^2
		&= 1 + \frac{2 + 1 - 2i\alpha\beta\sin(k) - 2\sqrt{2}\alpha\cos\left(\pm\omega_k-k\right) + 2\sqrt{2}i\beta\sin\left(\pm\omega_k\right)}{1+2i\alpha\beta\sin(k)} \\
		&= \frac{4 - 2\sqrt{2}\alpha\cos\left(\pm\omega_k-k\right) \pm 2\sqrt{2}i\beta\sin\left(\omega_k\right)}{1+2i\alpha\beta\sin(k)} .
\end{align*}
Thus the normalized eigenvectors of $M_k$ are
\[ \ket{\psi_\pm} = \sqrt{ \frac{1+2i\alpha\beta\sin(k)}{4 - 2\sqrt{2}\alpha\cos\left(\pm\omega_k-k\right) \pm 2\sqrt{2}i\beta\sin\left(\omega_k\right)} }
\begin{pmatrix} 1 \\ \frac{\alpha + \beta e^{-ik} - \sqrt{2} e^{i(\pm\omega_k - k)}}{\alpha - \beta e^{-ik} } \end{pmatrix}. \]
Using these eigenvalues and eigenvectors, we diagonalize $M_k$ and take matrix powers of it:
\[ \left( M_k \right)^t = \left(\lambda_+\right)^t \ketbra{\psi_+}{\psi_+} + \left(\lambda_-\right)^t \ketbra{\psi_-}{\psi_-}. \]
Then acting by this on the initial momentum wavefunction $\tilde\psi(k,0) = (1,0)^\intercal$, we get the momentum wavefunction at time $t$:
\begin{align*}
	&\tilde{\psi}(k,t)
	= e^{i\omega_kt} \frac{1+2i\alpha\beta\sin(k)}{4 - 2\sqrt{2}\alpha\cos\left(\omega_k-k\right) + 2\sqrt{2}i\beta\sin\left(\omega_k\right)} \begin{pmatrix} 1 \\ \frac{\alpha + \beta e^{-ik} - \sqrt{2} e^{i (\omega_k - k)}}{\alpha - \beta e^{-ik} } \end{pmatrix} \\
	&\quad+ e^{-i\omega_kt} \frac{1+2i\alpha\beta\sin(k)}{4 - 2\sqrt{2}\alpha\cos\left(\omega_k+k\right) - 2\sqrt{2}i\beta\sin\left(\omega_k\right)} \begin{pmatrix} 1 \\ \frac{\alpha + \beta e^{-ik} - \sqrt{2} e^{i (-\omega_k - k)}}{\alpha - \beta e^{-ik} } \end{pmatrix} .
\end{align*}
From this, we identify the left and right components:
\begin{align*}
	\tilde{\psi}_L(k,t)
		&= e^{i\omega_kt} \frac{1+2i\alpha\beta\sin(k)}{4 - 2\sqrt{2}\alpha\cos\left(\omega_k-k\right) + 2\sqrt{2}i\beta\sin\left(\omega_k\right)} \\
		&\quad+ e^{-i\omega_kt} \frac{1+2i\alpha\beta\sin(k)}{4 - 2\sqrt{2}\alpha\cos\left(\omega_k+k\right) - 2\sqrt{2}i\beta\sin\left(\omega_k\right)} 
\end{align*}
\begin{align*}
	&\tilde{\psi}_R(k,t) \\
		&\quad= e^{i\omega_kt} \frac{1+2i\alpha\beta\sin(k)}{4 - 2\sqrt{2}\alpha\cos\left(\omega_k-k\right) + 2\sqrt{2}i\beta\sin\left(\omega_k\right)} \frac{\alpha + \beta e^{-ik} - \sqrt{2} e^{i (\omega_k - k)}}{\alpha - \beta e^{-ik} } \\
		&\quad+ e^{-i\omega_kt} \frac{1+2i\alpha\beta\sin(k)}{4 - 2\sqrt{2}\alpha\cos\left(\omega_k+k\right) - 2\sqrt{2}i\beta\sin\left(\omega_k\right)} \frac{\alpha + \beta e^{-ik} - \sqrt{2} e^{i (-\omega_k - k)}}{\alpha - \beta e^{-ik} } .
\end{align*}
These are the left and right components of the momentum space wavefunctions.

\begin{figure}
\begin{center}
	\includegraphics{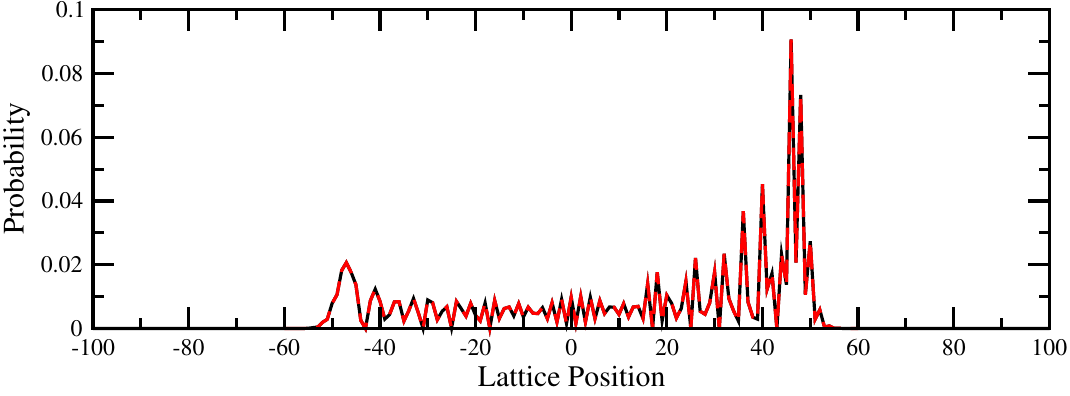}
	\caption{\label{fig:prob_time_integrate} Probability at each lattice site for an initially left-moving particle localized at 0 after 100 steps of the quantum walk with $\alpha = \cos(0.8)$ and $\beta = i\sin(0.8)$. The solid black curve is from iteratively simulating the evolution, and the dashed red curve is from integrating \eqref{eq:wavefunction_L} and \eqref{eq:wavefunction_R}. The two methods are in agreement, so the curves overlap.}
\end{center}
\end{figure}

To go to position space, we take the inverse Fourier transform of the momentum wavefunction:
\[ \psi(n,t) = \int_{-\pi}^\pi \frac{dk}{2\pi} \tilde\psi(k,t) e^{-ikn}. \]
Let us begin with the left direction. The inverse Fourier transform is
\begin{align*}
	\psi_L(n,t)
	&= \int_{-\pi}^\pi \frac{dk}{2\pi} e^{i\omega_kt} \frac{1+2i\alpha\beta\sin(k)}{4 - 2\sqrt{2}\alpha\cos\left(\omega_k-k\right) + 2\sqrt{2}i\beta\sin\left(\omega_k\right)} e^{-ikn} \\
	&\quad+ \int_{-\pi}^\pi \frac{dk}{2\pi} e^{-i\omega_kt} \frac{1+2i\alpha\beta\sin(k)}{4 - 2\sqrt{2}\alpha\cos\left(\omega_k+k\right) - 2\sqrt{2}i\beta\sin\left(\omega_k\right)} e^{-ikn}.
\end{align*}
Let's focus on the first integral:
\[ \int_{-\pi}^\pi \frac{dk}{2\pi} e^{i\omega_kt} \frac{1+2i\alpha\beta\sin(k)}{4 - 2\sqrt{2}\alpha\cos\left(\omega_k-k\right) + 2\sqrt{2}i\beta\sin\left(\omega_k\right)} e^{-ikn}. \]
Let $k = u + \pi$. Then this integral becomes
\[ \int_{-2\pi}^0 \frac{du}{2\pi} e^{i\omega_{u+\pi}t} \frac{1+2i\alpha\beta\sin(u + \pi)}{4 - 2\sqrt{2}\alpha\cos\left(\omega_{u+\pi}-(u+\pi)\right) + 2\sqrt{2}i\beta\sin\left(\omega_{u+\pi}\right)} e^{-i(u+\pi)n}, \]
where from \eqref{eq:omega},
\[ \cos\omega_{u+\pi} = \frac{\cos(u+\pi)}{\sqrt{2}} = \frac{-\cos{u}}{\sqrt{2}} = -\cos{\omega_u} \]
implies that
\[ \omega_{u+\pi} = \cos^{-1} (-\cos \omega_u) = \pi - \omega_u. \]
So this integral simplifies to
\[ \int_{-2\pi}^0 \frac{du}{2\pi} (-1)^{n+t} \frac{1-2i\alpha\beta\sin(u)}{4 - 2\sqrt{2}\alpha\cos\left(\omega_u+u\right) + 2\sqrt{2}i\beta\sin\left(\omega_u\right)} e^{-i(\omega_ut+un)}. \]
Since the integrand is periodic with period $2\pi$, we can shift the bounds to
\[ \int_{-\pi}^\pi \frac{du}{2\pi} (-1)^{n+t} \frac{1-2i\alpha\beta\sin(u)}{4 - 2\sqrt{2}\alpha\cos\left(\omega_u+u\right) + 2\sqrt{2}i\beta\sin\left(\omega_u\right)} e^{-i(\omega_ut+un)}. \]
Combining this with the second integral, the position-space wavefunction in the left direction is
\begin{align}
	\psi_L(n,t)
	&= \int_{-\pi}^\pi \frac{dk}{2\pi} (-1)^{n+t} \frac{1-2i\alpha\beta\sin(k)}{4 - 2\sqrt{2}\alpha\cos\left(\omega_k+k\right) + 2\sqrt{2}i\beta\sin\left(\omega_k\right)} e^{-i(\omega_kt+kn)} \notag \\
	&\quad+ \int_{-\pi}^\pi \frac{dk}{2\pi} \frac{1+2i\alpha\beta\sin(k)}{4 - 2\sqrt{2}\alpha\cos\left(\omega_k+k\right) - 2\sqrt{2}i\beta\sin\left(\omega_k\right)} e^{-i(\omega_kt+kn)}. \label{eq:wavefunction_L}
\end{align}
Similarly, the right-pointing wavefunction is
\begin{align}
	\psi_R(n,t)
	&= \int_{-\pi}^\pi \frac{dk}{2\pi} (-1)^{n+t} \frac{1-2i\alpha\beta\sin(k)}{4 - 2\sqrt{2}\alpha\cos\left(\omega_k+k\right) + 2\sqrt{2}i\beta\sin\left(\omega_k\right)} \notag \\ &\quad\quad\quad\quad\quad\times \frac{\alpha - \beta e^{-ik} - \sqrt{2} e^{-i (\omega_k + k)}}{\alpha + \beta e^{-ik} } e^{-i(\omega_kt+kn)} \notag \\
	&\quad+ \int_{-\pi}^\pi \frac{dk}{2\pi} \frac{1+2i\alpha\beta\sin(k)}{4 - 2\sqrt{2}\alpha\cos\left(\omega_k+k\right) - 2\sqrt{2}i\beta\sin\left(\omega_k\right)} \notag \\ &\quad\quad\quad\quad\quad\quad\times \frac{\alpha + \beta e^{-ik} - \sqrt{2} e^{-i (\omega_k + k)}}{\alpha - \beta e^{-ik} } e^{-i(\omega_kt+kn)}. \label{eq:wavefunction_R}
\end{align}
Although these formulas \eqref{eq:wavefunction_L} and \eqref{eq:wavefunction_R} are complicated, they can be numerically integrated and verified to agree with a numerical simulation of the quantum walk; this is shown in Figure~\ref{fig:prob_time_integrate}, where the two probability distributions generated by each method overlap in agreement. Also note that the wavefunction at every other lattice site is no longer equal to zero with the potential barriers, even though they were without the barriers.

\begin{figure}
\begin{center}
	\includegraphics{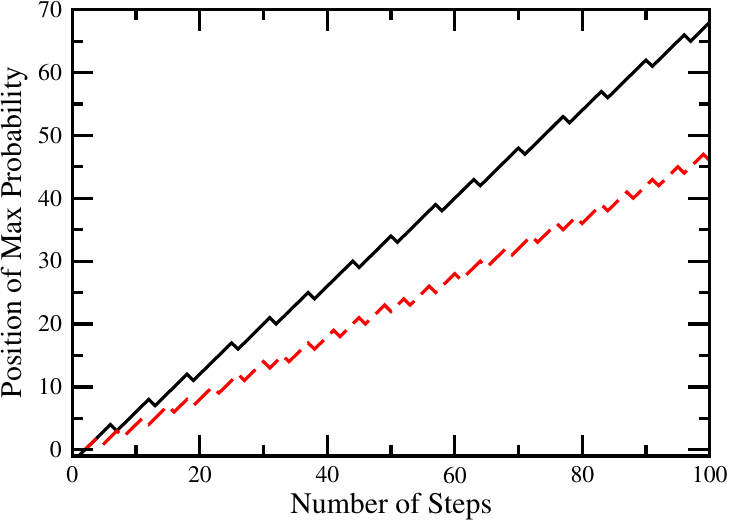}
	\caption{\label{fig:peaks} Lattice position of the maximum probability as steps of the quantum walk \eqref{eq:U} are taken with an initially left-moving particle localized at 0. The solid black curve is with $\alpha = 1$ and $\beta = 0$, and the dashed red curve is with $\alpha = \cos(0.8)$ and $\beta = i\sin(0.8)$.}
\end{center}
\end{figure}

Now that we have the position-space wavefunctions in each direction, we can approximate them for large times ($t \gg 1$). As described in \cite{NV2000}, this can be done using the method of steepest descent, where we examine the time-oscillatory portion of \eqref{eq:wavefunction_L} and \eqref{eq:wavefunction_R}, which is
\[ e^{-i(\omega_kt+kn)} = e^{-i \phi(k,\alpha) t}, \]
where we have defined
\[ \phi(k,\alpha) = \omega_k + \frac{kn}{t}. \]
From the analysis of \cite{NV2000}, the part we care about is the movement of the peak in probability (\textit{e.g.}, the leading right peak in Figure~\ref{fig:prob_time_barrier}). This is given by where $\phi$ has first and second derivative (with respect to $k$) equal to zero. So let us calculate these derivatives, beginning with the first:
\[ \frac{\partial \phi}{\partial k} = \omega_k' + \frac{n}{t}. \]
To find $\omega_k'$, we use implicit differentiation on $\omega_j$'s definition \eqref{eq:omega}, which yields
\[ \omega_k' = \frac{\alpha\sin k}{\sqrt{2-\alpha^2\cos^2k}}. \]
Plugging this in, the first derivative of $\phi$ is
\[ \frac{\partial \phi}{\partial k} = \frac{\alpha\sin k}{\sqrt{2-\alpha^2\cos^2k}} + \frac{n}{t}. \]
Now let us differentiate this to find the second derivative of $\phi$:
\[ \frac{\partial^2 \phi}{\partial k^2} = \frac{\alpha \cos k \left( 2 - \alpha^2 \right)}{\left( 2 - \alpha^2 \cos^2k \right)^{3/2}}. \]
Using these expressions for the first and second derivative of $\phi$, we find they are both zero when
\[ k = \frac{\pi}{2}, \frac{3\pi}{2} \quad \text{and} \quad \frac{n}{t} = \mp \frac{\alpha}{\sqrt{2}}. \]
It is easily verified that this is a 2nd order stationary point, so $\partial^3 \phi / \partial k^3 \ne 0$ at these values. Thus the peak occurs at position
\[ n = \pm \frac{\alpha}{\sqrt{2}} t \]
at time $t$, which is the quantum walk's characteristic linear dispersion $\mathrm{\Theta}(t)$. That is, the spreading remains ballistic, with the potential barriers only affecting the constant factor. Note when $\alpha = 1$, we get the expected potential-free peak at $t/\sqrt{2}$. This is confirmed in Figure~\ref{fig:peaks}, where we plot the location of the peak as the quantum walk operator \eqref{eq:U} is applied; the potential-free curve has a linear regression with slope nearly $1 / \sqrt{2} \approx 0.707$, and with $\alpha = \cos(0.8)$, the slope is nearly $\cos(0.8)/\sqrt{2} \approx 0.493$.

This result can also be used to detect and quantify the hopping errors in experiments. If the movement of the quantum walk has a speed slower than $1/\sqrt{2}$ lattice positions per unit time, then one might suspect that hopping errors are occurring of the form $(\alpha S + \beta I)$. Since the speed equals $\alpha / \sqrt{2}$, this in turn yields $\alpha$ and $\beta = i \sqrt{1 - \alpha^2}$.

%-------------------------------------------------------------------------------
% Section
%-------------------------------------------------------------------------------

\section{Continuous-Time Quantum Walks}

We end with a short comment on continuous-time quantum walks \cite{FG1998b}, which evolve on $k$-regular graphs by Schr\"odinger's equation with Hamiltonian
\[ H = -\gamma A, \]
where $\gamma$ is the jumping rate (\textit{i.e.}, amplitude per time) of the walk, and $A$ is the adjacency matrix of the graph ($A_{ij} = 1$ if vertices $i$ and $j$ are connected and $0$ otherwise). The potential barriers would hinder the amplitude from jumping, say by $\epsilon$, and cause it to stay put instead. This effectively causes the adjacency matrix to be \cite{Wong2015c}
\[ A' = k \epsilon I + (1-\epsilon) A. \]
But the multiple of the identity matrix can be dropped since it only contributes an overall, unobservable phase. Thus the system effectively evolves with Hamiltonian
\[ H' = -\gamma(1-\epsilon)A = (1-\epsilon) H. \]
But this is just a rescaling of the potential-free Hamiltonian, so the system evolves the same way, but with rescaled time $t' = (1-\epsilon) t$. So assuming that the potential barrier yields a reduction in amplitude $\epsilon$ that does not approach $1$, the ballistic transport of the continuous-time quantum walk \cite{Childs2003} is retained, similar to the discrete-time quantum walk we showed above.

%-------------------------------------------------------------------------------
% Section
%-------------------------------------------------------------------------------

\section{Conclusion}

We have analyzed the discrete-time quantum walk on the infinite 1D lattice, replacing the typical moving shift with the flip-flop shift, which is the one actually used in quantum walk search algorithms. With this, we considered the effect of potential barriers hindering the hop, showing that as long as the barriers do not asymptotically cause the walk to stop, they do not change the $\mathrm{\Theta}(t)$ ballistic transport of the quantum walk. This does change the coefficient of the dispersion, however, and this provides a way to detect and quantify the hopping errors. These results also apply to continuous-time quantum walks.

Since our model assumes that the potential barriers are equal everywhere on the lattice and do not fluctuate, an open question is the effect of nonhomogeneous or random barriers. This contrasts with previous work on nonhomogeneous or random coins \cite{Joye2010,Ahlbrecht2011}, and comparing or combining the investigations would be particularly interesting.

%-------------------------------------------------------------------------------
% Acknowledgments
%-------------------------------------------------------------------------------

\begin{acknowledgements}
	Thanks to Andris Ambainis and Alexander Rivosh for useful discussions. This work was partially supported by the European Union Seventh Framework Programme (FP7/2007-2013) under the QALGO (Grant Agreement No.~600700) project, and the ERC Advanced Grant MQC. 
\end{acknowledgements}

%-------------------------------------------------------------------------------
% References
%-------------------------------------------------------------------------------

\bibliographystyle{qinp}
\bibliography{refs}

\end{document}